\documentclass[preprint,12pt,showkeys]{revtex4}
\usepackage{ams}
\usepackage{graphicx}

\usepackage{color}

\usepackage{fancyhdr}
\begin{document}

\title{Switching Mechanism in Ferromagnetic Nanorings }


\author{Wen Zhang}
\email{zhangwen@usc.edu}
\author {Stephan Haas}
\email{shaas@usc.edu}

\affiliation{Department of Physics and Astronomy, University of
Southern California, Los Angeles, CA 90089, USA\\}

\begin{abstract}
Ferromagnetic rings exhibit many novel physical phenomena with
promise for potential applications. Here, we focus on switching
processes which are fundamental properties of magnetic systems and
are especially crucial for data storage applications. A brief
introduction on the advantages of ring geometries is presented,
followed by a background survey of the model and methods to study
nanomagnetic elements. The relevant magnetic states identified until
now are discussed, with special emphasis given to vortex, onion, and
twisted states. Such states are potential candidates for data
storage applications. Switching processes under uniform fields are
categorized into three types, and their mechanisms are explained in
detail. In particular, explanations from a topological point of view
are shown to be enlightening. We conclude with a brief discursion of
circular field switching and data storage applications of
magnetic nanorings.\\


\end{abstract}

\maketitle

\tableofcontents

\clearpage
\section{Introduction}
During the last two decades, tremendous effort has been devoted to
micron and submicron magnetic elements, motivated both by
fundamental scientific interest, such as novel magnetic states and
switching processes \cite{De'Bell,Liu,Cowburnreview}, and by their
potential for technological applications, such as magnetic random
access memory (MRAM)  \cite{cmuJAP2000,MRAM} and magnetic sensors
\cite{sensor1,sensor2}. Characterizing the magnetic properties of
nanostructures is a challenging task, as their shape significantly
influences their physical response. Significant work has been
invested into identifying the geometries which offer the simplest,
fastest, and most reproducible switching mechanisms. Particular
attention has focused on magnetic structures with high-symmetry
geometries, such as circular disks and squares, since spin
configurations with high symmetry are expected for these elements,
which in turn are believed to yield simple and reproducible memory
states.

\begin{figure}[h]\begin{center}
\includegraphics[height=.3\textheight]{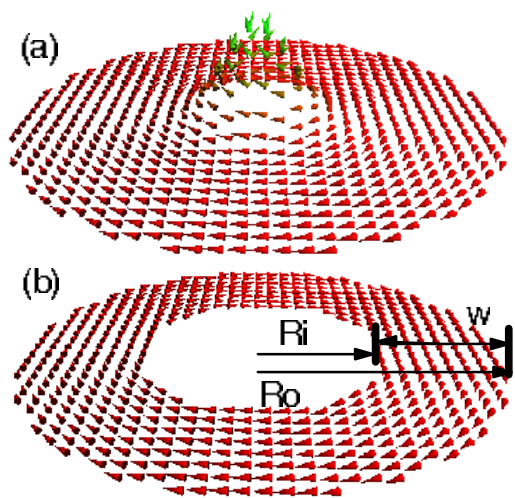}
\caption{\label{vortex}(color online) Schematic representation of a
magnetic vortex state in a disc (a) and a ring (b). The core region
where spin points out of plane are exaggerated and represented by
the height and green color. $R_o$ and $R_i$ are outer and inner
radius of rings and w is their width. }
\end{center}\end{figure}

In analogy to the traditional approach to encode information in
dipolar-like giant spins, quasi-uniform single domain states have
been proposed and intensively studied. However, these typically
suffer from three fundamental disadvantages: (i) they are sensitive
to edge roughness so that the switching field typically has a broad
distribution; (ii) because of the long-ranged dipolar interactions
between separate elements, high density arrays are hard to achieve;
(iii) they cannot be made to be too small due to the
superparamagnetism effect. To overcome these problems, use of the
magnetic vortex state in disc geometries (see Fig. \ref{vortex}(a))
has been suggested, since it is insensitive to edge imperfections
and entirely avoids the superparamagnetism effect. Moreover, the
zero in-plane stray field opens the possibility of high density
storage. Nevertheless, the vortex state is stable only in discs of
fairly large sizes (diameter over about 100nm) due to the existence
of high energy penalty vortex core regions. What is worse, the
switching mechanisms for discs are complex and difficult to control.
In seeking a solution, ring geometries (see Fig. \ref{vortex}(b))
have been proposed and studied intensively in the last decade. In
addition to all the advantages of vortex states in the disc
geometry, magnetic rings are completely stray field free and can be
stable with diameters as small as 10nm. Taking 10nm spacing into
account, ring arrays give an ultimate area storage density of about
$0.25 Tbits/cm^2$ (or $1.6Tbits/in.^2$), substantially higher than
the traditional hard disc area storage density limit. Besides this
promising application potential, ring geometries have proven to be a
wonderful platform for the investigation of fundamental physical
questions concerning domain walls. \cite{KlauiReview2008}

For the study of magnetic properties of structures in micrometer and
nanometer dimensions, micromagnetic simulations prove to be a very
powerful tool. \cite{simulationreview} To study dynamic properties,
one has to resort to numerical integration of the
Landau-Lifshitz-Gilbert (LLG) equation \cite{ll,gilbert}. For
(quasi)static properties, Monte Carlo simulations are also applied
extensively. \cite{De'Bell} Numerically evaluating the energy of
such systems is highly nontrivial because of the long-range
character of the dipolar interaction between magnetic moments which
contributes to the magnetostatic energy. The complexity of a
brute-force calculation is $\mathcal{O}$($N^2$). However, there are
several approaches to improve this. The most important two
techniques are the Fast Fourier Transform (FFT), which reduces the
complexity to $\mathcal{O}$($Nlog(N)$), and the Fast Multipole
Method (FMM) \cite{greengard, me2}, which reduces the complexity to
$\mathcal{O}$($N$) but with fairly large coefficients. Based on
these methods, several open software packages are available:
OOMMF\cite{oommf}, magpar\cite{magpar}, Nmag\cite{Nmag}, $\psi$-mag
\cite{psimag}. Of these, OOMMF is the most mature and popular one,
and has been extensively used worldwide.


This review article is an overview of current research on magnetic
rings, with a focus on clarifying their quasistatic properties. We
will mainly refer to results in the literature and include some of
our own recent findings. Considering the overwhelming amount of
literature, we apologize for the inevitable omission of some
important works. The article is organized as follows: in section 2,
we cover the basic background needed to study magnetic systems; in
section 3, we explain the equilibrium magnetic states (spin
configuration) and metastable states in ring geometries; finally,
the mechanism of switching processes in rings is discussed in detail
in section 4, followed by a brief discussion of data storage
applications in section 5.

\section{Background}
In the quasi-classical approximation, the Hamiltonian
($\mathcal{H}$) of a magnetic nanoparticle in a magnetic field
consists of four terms: exchange interaction, dipolar interaction,
crystalline anisotropy and Zeeman energy. The sum of the first three
terms yields the internal energy. If each magnetic moment occupies a
site of the underlying crystal lattice, $\mathcal{H}$ is given by

\begin{eqnarray}\label{sum}
\mathcal{H}&=&-J\sum_{<i,j>}\vec{S}_i\cdot\vec{S}_j
+D\sum_{i,j}\frac{\vec{S}_i\cdot\vec{S}_j-3(\vec{S}_i\cdot
\hat{r}_{ij})(\vec{S}_j\cdot \hat{r}_{ij})}{r_{ij}^{3}} +U_k -
\vec{H}\cdot \sum_i \vec{S}_i ,
\end{eqnarray}

where $J>0$ is the ferromagnetic exchange constant (or exchange
integral, measured in units of energy), which is assumed to be
non-zero only for nearest neighbors, $D$ is the dipolar coupling
parameter and $\vec{r}_{ij}$ is the displacement vector between
sites $i$ and $j$. The anisotropy term $U_k$ can take various
forms\cite{Kittel}, among which the most common are uniaxial
anisotropy $U_k=K\sum_i sin^2\theta_i$ , where $\theta_i$ is the
angle $\vec{S}_i$ makes with the easy axis, and cubic anisotropy
$U_k=K\sum_{i}[\alpha_i^2 \beta_i^2+\beta_i^2 \gamma_i^2+\alpha_i^2
\gamma_i^2]$, where {$\alpha_i,\beta_i,\gamma_i$} are the direction
cosines of $\vec{S}_i$. Note that $K$ is the single site anisotropy
energy (not an energy density) and that $\vec{S}$ is a dimensionless
unit spin vector with magnetic moments $\vec\mu=|\mu|\vec{S}$.
Experimentally, the materials used are mostly Co, Fe, permalloy and
supermalloy. For these materials, the ratio $D/Ja^3$ falls in the
range of $10^{-3}$ and $10^{-5}$, where $a\sim0.3nm$ is lattice
constant. For polycrystalline systems, $U_k$ is usually omitted
since the crystalline anisotropy is very small. For epitaxial
systems, however, crystalline anisotropy needs to be considered.
Generally it stabilizes certain directions and enhances the
switching field. It also helps to resist the superparamagnetism
effect. The Zeeman term above applies for uniform fields. If this is
not the case, the term is adjusted to $ \sum_i
\vec{S}_i\cdot\vec{H}_i$. The main observables in magnetic systems
are the magnetization and susceptibility.

The competition between the various energy terms in $\mathcal{H}$
results in the interesting complexity of magnetic nanostructures.
The exchange interaction tends to align spins in the same direction,
whereas the dipolar interaction encourages spins to line up along
the boundaries, which is the origin of the shape anisotropy. Thus
spins align in-plane in a flat element while they point out-of-plane
if the thickness is much larger than the lateral dimension. The
former case attracts more attention since it exhibits rich new
phenomena and is closely related to applications. Plenty of
competing configurations exist, including flower, leaf, onion,
buckle, and vortex states.

The study of micron and submicron magnetic elements has two major
branches: quasistatic properties and dynamics. Here we will focus on
the former, in particular on the switching process, a first order
phase transition. To understand the switching process, we first
discuss in detail all the possible states in magnetic rings. After
that, the magnetization reversal mechanism (the hysteresis) is
investigated.

\begin{figure}[h]
\begin{center}
\includegraphics[height=.4\textheight]{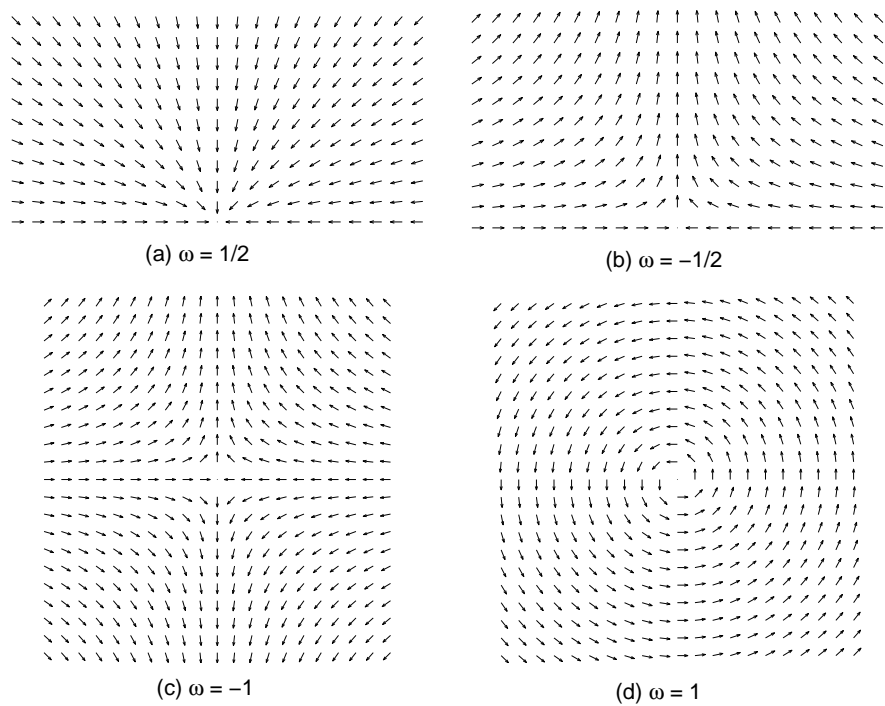}
\caption{\label{tdefect} Elementary topological defects: (a) edge
defect $\omega$=1/2, (b) edge defect $\omega$=-1/2, (c) vortex
$\omega$=1, (d) antivortex $\omega$=-1.}

\end{center}
\end{figure}

Many magnetic phenomena involve vortices and domain walls. From a
topological point of view, these are defects. The topological theory
of defects  \cite{top79,top2005,top2006} can help us to understand
the complex creation and annihilation processes during switching.
One of the most important principles is the conservation of
topological charge. Topological charge is defined as the winding
number $\omega$, i.e. a line integral around the defect center:

\begin{eqnarray}\label{w}
\omega=\frac{1}{2\pi}\oint \nabla \theta(\vec{r}) \cdot d\vec{r},
\end{eqnarray}

where $\theta$ is the angle between the local magnetic moment and
the positive x axis. There are several elementary topological
defects in magnetic systems. As is known, a vortex in the bulk has
an integer winding number $\omega=1$ and an antivortex has
$\omega=-1$. Furthermore, Tchernyshyov \textsl{et al} \cite{top2005}
identified two edge defects with fractional winding numbers
$\omega=\pm1/2$. These four types of elementary topological defects
are shown in Fig.\ref{tdefect}. All the magnetic intricate textures,
including domain walls, are composite objects made of some of the
above four elementary defects.\cite{top2005} These defects have
several properties: (i) in analogy to Coulomb interaction, two
defects with the same winding number sign repel each other, while
they attract each other if they have opposite signs of winding
number; (ii) the vortex is repelled by the boundary since it has an
image ``charge" with the same sign; (iii) edge defects are confined
to the edge by an effective confining potential; (iv) direct
annihilation of two defects with the same sign is prohibited; (v)
edge defects can change sign by introducing a bulk defect; (vi) in
sufficiently narrow rings, $\omega=\pm1/2$ edge defects are
degenerate, i.e. they have the same energy, whereas in thick rings
the $\omega=+1/2$ will have substantially higher energy than the
$\omega=-1/2$ so that it can decay into a vortex $\omega=1$ and an
edge defect $\omega=-1/2$.

Concerning experimental fabrication techniques, interested readers
can refer to Martin \textsl{et al} \cite{Liu} and Kl$\ddot{a}$ui
\textsl{et al} \cite{KlauiReview2003}. Experimentalists are able to
produce two kinds of morphology: polycrystalline and epitaxial. In
most cases, magnetic rings are made of polycrystalline permalloy,
supermalloy and Co, whose crystalline anisotropy is negligible. On
the other hand, when one wants to study the effects of anisotropy
and underlying lattice structure, epitaxial face centered cubic
(fcc), face centered tetragonal (fct) and hexagonal close-packed
(hcp) structure Co rings can be produced. The imaging techniques can
be classified into two groups: (i) intrusive techniques such as
magnetic force microscope (MFM): this method affects the magnetic
state of the sample because of the external magnetic field of the
tip; (ii) non-intrusive techniques such as Photoemission Electron
Microscopes (PEEM) and Scanning Electron Microscopy with
Polarization Analysis (SEMPA): the magnetic states remain the same
after the image is scanned. The spatial resolution nowadays can be
as high as 10nm. Atomic resolution is also reported by SP-STM
\cite{spstm, spstm2}. To measure the hysteresis curve, the most
popular techniques used are the magneto-optic Kerr effect (MOKE) and
SQUID. Other techniques include measurement of the Hall
resistance\cite{triple} and of the spin-wave
spectrum\cite{spinwave1}. In most cases, the hysteresis is measured
in arrays of nanorings, typically larger than $10\times10$. Thus the
transition field has a range which represents the transition field
distribution among these nanoparticles. A few measurements have been
reported for single particles where the transition is sharp, but the
size of these particles is fairly big.



\section{Magnetic States}

A first step in the investigation of the magnetic properties of
certain geometries is to identify the competing stable and
metastable spin configurations at remanence. This information can be
summarized in a geometric phase diagram.

\begin{figure}[h]\begin{center}
\includegraphics[height=.3\textheight]{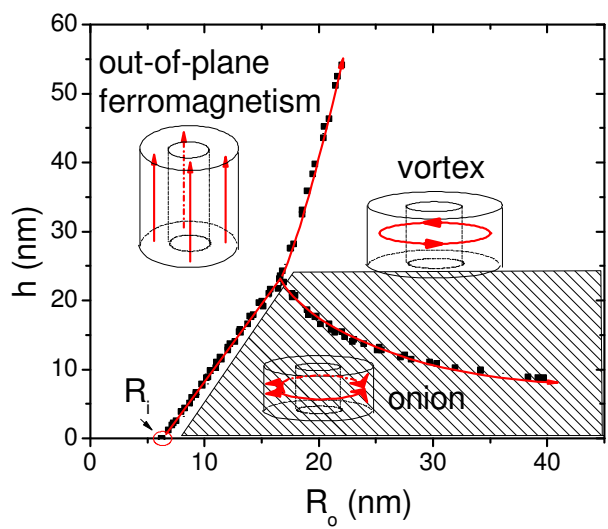}
\caption{\label{rp}(color online) Schematic geometric magnetic phase
diagram for nanorings with $R_i=6nm$. Shaded area is the region
where the onion state could be (meta)stable, and thus it is of
greatest interest.}

\end{center}\end{figure}

Several works have been devoted to determining this phase diagram
\cite{nanoring1, nanoring2, me} for ring structures.  Fig. \ref{rp}
shows an example with inner radius $R_i$ fixed. Spins point out of
plane in parallel when the element height $h$ is much larger than
width $w$. The opposite situation is of more interest. When $h/w<2$,
there are two possible states: (i) the vortex state (V) (or the flux
close state), characterized by the circulation of spins; (ii) the
onion state (O) (or the quasi uniform state), characterized by two
head-to-head domain walls. The vortex state is the only single
domain state in ring structures. Because of the resemblance to
strips, magnetic rings hold a lot of multi-domain states, such as
onion states and various twisted states discussed in detail below.
The onion state has two domains, separated by two head-to-head
domain walls. The two domains are sometimes referred to two arms.
Since the domain walls in ring geometries are well confined, ring
magnets serve as a perfect platform to study the motion of domain
walls as well as the interaction between them
\cite{KlauiReview2003,KlauiReview2008}.
%

\begin{figure}[h]\begin{center}
\includegraphics[height=.5\textheight]{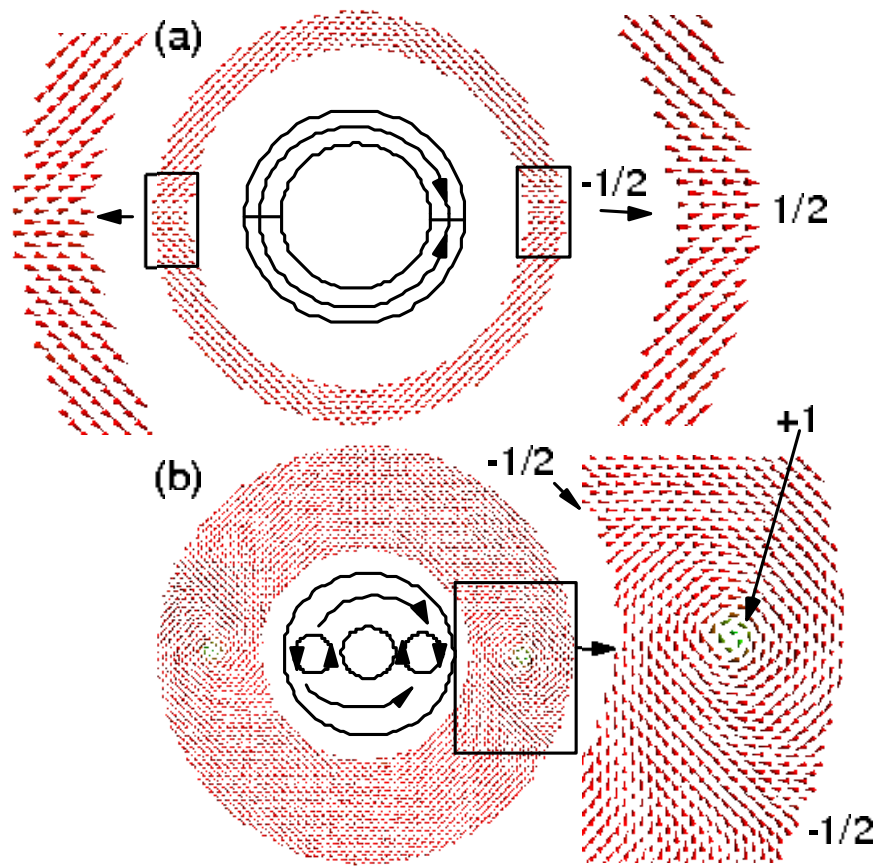}
\caption{\label{domainwall}(color online) Head-to-head domain wall.
(a) onion state with transverse domain wall; (b) onion state with
vortex domain wall. The numbers in the figure indicate the type of
the topological defect. }

\end{center}\end{figure}

\begin{figure}[h]\begin{center}
\includegraphics[height=.3\textheight]{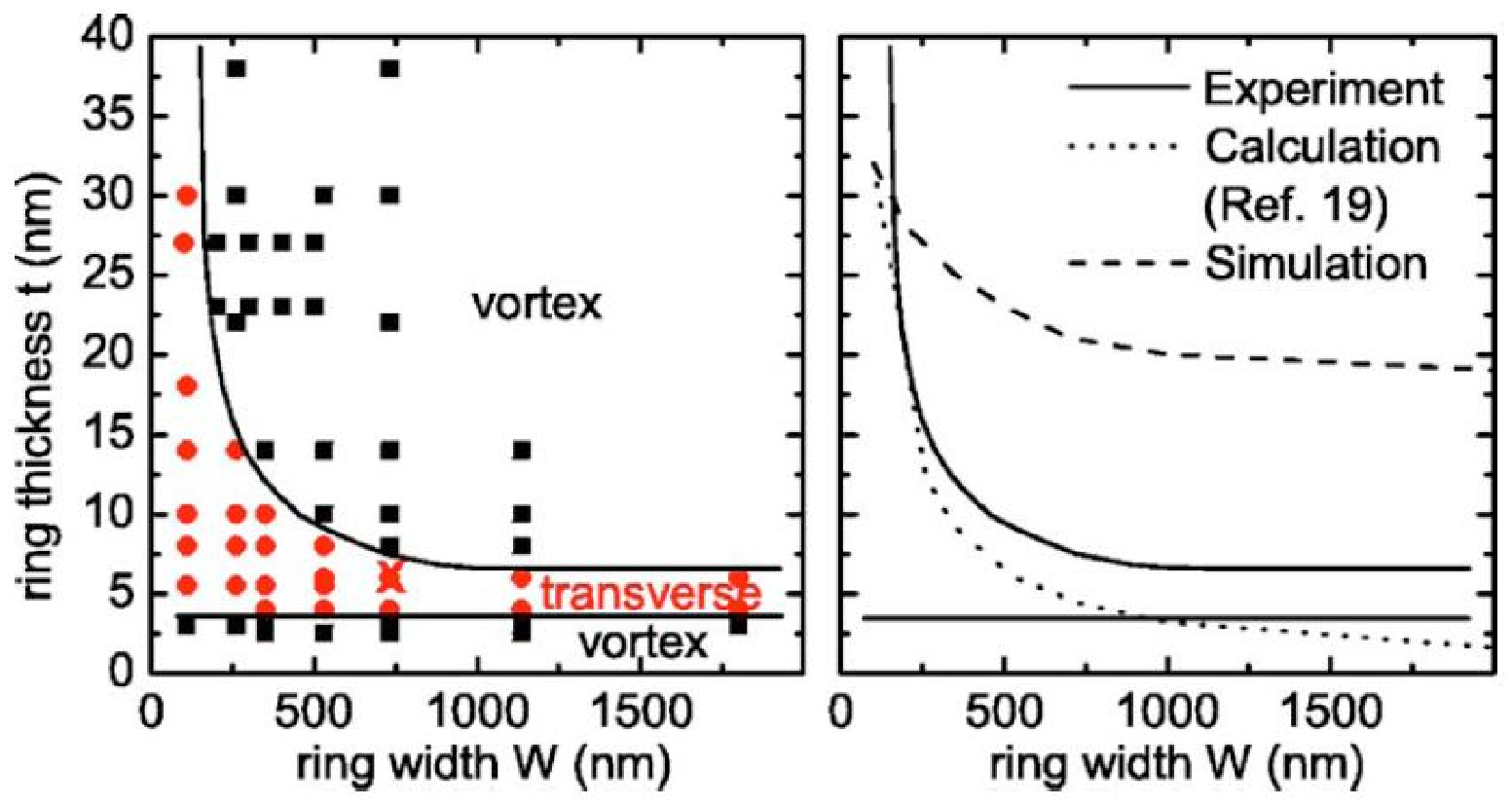}
\caption{\label{wallp} \cite{KlauiAPL2006}(\copyright{Reused with
permission from M. Laufenberg, Applied Physics Letters, 88, 052507
(2006). Copyright 2006, American Institute of Physics.}) (color
online) ``(a) Experimental phase diagram for head-to-head domain
walls in NiFe rings at room temperature. Black squares indicate
vortex walls and red circles transverse walls. The phase boundaries
are shown as solid lines. (b) A comparison of the upper experimental
phase boundary (solid line) with results from calculations (dotted
line) and micromagnetic simulations (dashed line). Close to the
phase boundaries, both wall types can be observed in nominally
identical samples due to slight geometrical variations. The
thermally activated wall transitions shown were observed for the
ring geometry marked with a red cross (W=730 nm, t=7 nm)."}
\end{center}\end{figure}

All the interesting behavior of magnetic systems stems from the fact
that there is a great variety of metastable states. Several states
may be stable at remanence depending on the history how remanence is
reached. One can see from Fig. \ref{rp} that the dominant phase for
thin films is the vortex state, but the onion state is actually a
metastable state in a very large region. If the remanence state is
obtained by relaxing from saturation, it is usually an onion state.
A remanent state phase diagram is given for Co nanorings
\cite{francePRL2001} where the onion state area is significantly
increased. As mentioned above, onion states are characterized by two
head-to-head domain walls\cite{hh} which have long been studied in
magnetic strips. Two kinds of head-to-head domain walls exist:
transverse domain walls in thin rings, see Fig. \ref{domainwall}(a)
and vortex walls in wide rings, see Fig. \ref{domainwall}(b). A
geometric phase diagram (see Fig. \ref{wallp}) for the two
head-to-head domain walls was found by Laufenberg et al
\cite{KlauiAPL2006}.

Note that the phase diagram was obtained by relaxing the system from
saturation, so it does not represent the ground state of the system
at zero field and shows no vortex state at all. One can see from the
Fig. \ref{wallp}(b) that analytical calculations tend to favor
vortex walls while simulations tend to favor transverse walls
compared with the experimental result. It results from the fact that
there is an energy barrier between vortex walls and transverse
walls. Analytical calculations give the lower energy state, while
real systems can stay in the local minimum with transverse walls.
Simulations are performed in zero temperature, so that it is harder
to form vortex walls than in the experimental situation, where
thermal fluctuation can help the system overcome the barrier.


From a topological point of view, the transverse wall is composed of
a $\omega=1/2$ defect at the outer edge and a $\omega=-1/2$ defect
at the inner edge\cite{top2005}. The vortex wall is composed of two
$\omega=-1/2$ defects respectively at the outer and inner edge
together with a $\omega=+1$ vortex defect at the center of the
rim.\cite{top2006}

%

\begin{figure}[h]\begin{center}
\includegraphics[height=.3\textheight]{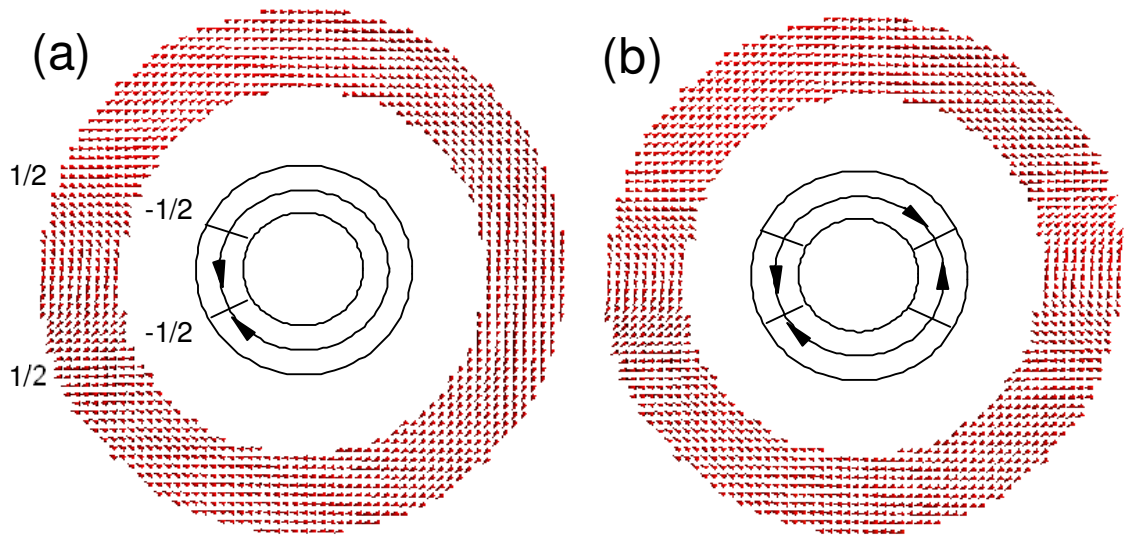}
\caption{\label{360d}(color online) Twisted states: (a) with single
$360^o$ domain wall, (b) with two $360^o$ domain walls. The numbers
in the figure indicate the type of defects.}

\end{center}\end{figure}

Recently a new category of metastable states has been
discovered\cite{RossPRB2003} in relatively small rings made of both
Co and permalloy with $R_o\sim180-520nm, w\sim30-200nm, h\sim10nm$:
the twisted states (T) (or saddle state\cite{saddle}). These states
are characterized by $360^o$ domain walls
\cite{360domain2,360domain} (see Fig. \ref{360d}), which have
previously been reported in narrow thin film strips.
\cite{twist1,twist2} They are shown to be stable within a field
range of several hundred Os \cite{RossPRB2003}. Furthermore, they
have low stray fields and are easy to control by current, which
makes them attractive for data storage applications. There can be
more than one $360^o$ domain wall in rings, resulting in a
multi-twisted state. The $360^o$ domain walls can be viewed as two
transverse domain walls (Fig. \ref{360d}(b)). The attraction between
the two transverse walls occurs because they have opposite senses of
rotation. This tendency is balanced by the exchange energy in the
region between the walls. The existence of this domain wall can also
be understood fairly easily by topological arguments, referring to
the properties (iv) and (vi) of topological defects in Sec. II. When
the width is small, the two transverse domain walls have the same
defect on the same side, and they are stable. So these edge defects
cannot annihilate, resulting in a $360^o$ domain wall. When the
width is larger, however, one of the $\omega=+1/2$ defects becomes
unstable and transforms into a $\omega=-1/2$ defect by introducing a
vortex into the center of the rim. Then the two transverse domain
walls are able to annihilate each other, resulting in a vortex
state. Therefore, only a portion of the region of stable transverse
wall in the phase diagram Fig. \ref{wallp} can possibly hold twisted
states. Twisted states in large rings have almost zero remanence, so
magnetization is incapable of distinguishing between vortex and
twisted states and other quantities must be used. Toroidal
moment\cite{toroidal} and winding number are supplementary order
parameter candidates.

In the presence of an external magnetic field, several other states
exist: wave states and shifted vortex core states. They usually
exist in very thick rings, which is of less interest for practical
reasons. Regarding the effect of crystalline anisotropy, it
stabilizes quasi uniform states (like onion states) and imposes some
additional domain structures.



\section{Switching Processes}

For data storage applications, understanding switching processes is
crucial. The main effort is directed towards identifying simple
reproducible switching processes and reliable sensitive detection
techniques. Meanwhile, in terms of fundamental physics, the
understanding of switching processes is just as important. As
picosecond and nanometer time and spatial resolution detection
techniques are still limited, the microscopic details of switching
processes and other dynamic transitions are currently mainly
investigated by micromagnetic simulations. One issue that is hard to
quantify is the distribution of the switching field.
Computationally, the transition in the hysteresis curve is very
sharp, while the experimental curves have a broad transition region
caused by edge roughness and defects.


The traditional way of studying switching processes is by applying a
uniform magnetic field with changing magnitude. Motivated by the
vortex configuration, circular fields are also studied in order to
switch the two vortex states with opposite circulation. Here we
mainly discuss the uniform field switching case and say a few words
about the circular field afterwards.

\begin{figure}[h]\begin{center}
\includegraphics[height=.3\textheight]{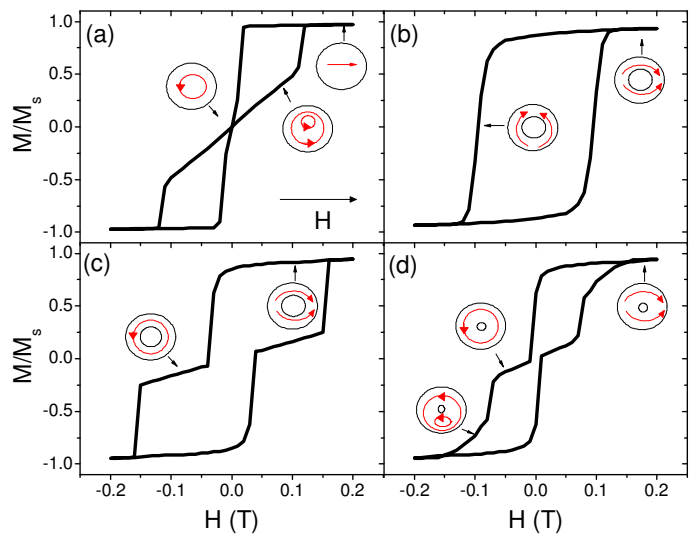}
\caption{\label{switching}(color online) Schematic switching process
and hysteresis: (a) hysteresis of magnetic discs, (b) one-step
switching of rings, (c) double switching of rings, (d) triple
switching of rings. }
\end{center}\end{figure}

\subsection{Uniform field}

With a uniform field, three types of typical hysteresis curves are
identified:

\subsubsection{One-step or single switching}
This is a direct onion-to-onion-state switching (O-O).
Fig.\ref{switching}(b)) shows a typical hysteresis curve. In this
process, an onion state is reversed to the opposite onion state
directly. This process is usually only observed experimentally in
small rings, especially in the thin film limit (small $h$).
Theoretically speaking, it should happen in much thicker system if
the ring is perfectly symmetric, as it is observed in simulations.
For real systems, however, there always exists some sort of
asymmetry caused either by defects or by the environment. When
asymmetry exists, the system tends to fall into its true ground
vortex state, which will result in a double switching process.

It is believed that the switching mechanism is simply a coherent
rotation, where two domain walls move in the same rotational
direction. This can easily be observed computationally if one
intentionally introduces some sort of asymmetry into the system.


\begin{figure}[h]\begin{center}
\includegraphics[height=.5\textheight]{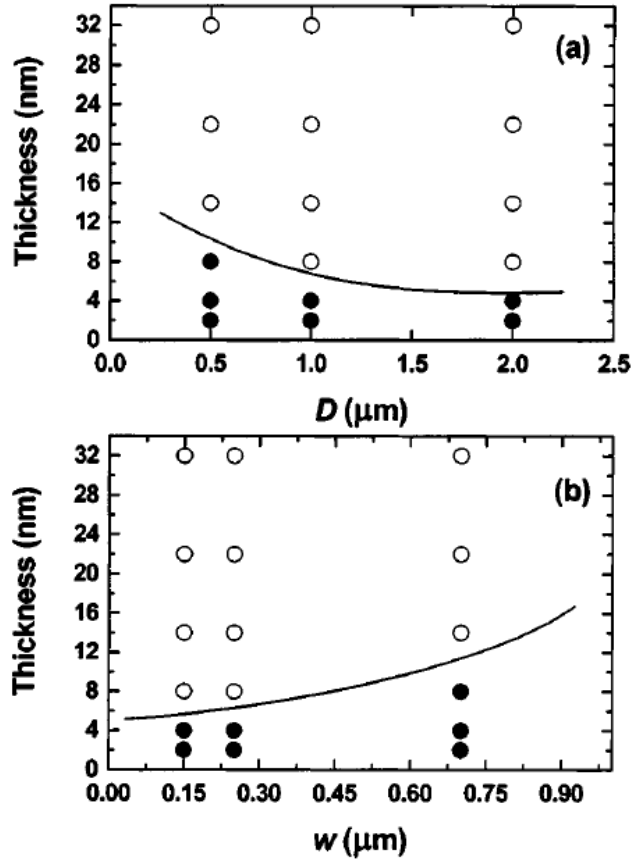}
\caption{\label{switchingp} \cite{KlauiAPL2003}(\copyright{Reused
with permission from Y. G. Yoo, Applied Physics Letters, 82, 2470
(2003). Copyright 2003, American Institute of Physics.}) ``Phase
diagrams of two-step switching (open circles) and single switching
(full circles) as a function the ring geometrical parameters. (a)
For a ring width of 0.25 mm. (b) For a ring diameter of 2 mm. The
solid lines define the boundary between the two different switching
regimes." }
\end{center}\end{figure}

Yoo \textsl{et al}\cite{KlauiAPL2003} offered experimental phase
diagrams for Co rings with constant width and $R_o$ (equivalent to
$D/2$ in Fig. \ref{switchingp}) and varying height $2\sim34nm$. They
conclude that the main geometric factor is the film thickness. When
$h<~6nm$, the above one-step switching happens, while otherwise it
is two-step switching. For small radii this is not necessarily true,
since they only study $R_o>500nm$. Rings with large radii resemble
straight strips very well. With shrinking radius, however, the story
is different. Since magnetic processes including nucleation and
annihilation depend strongly on the curvature, small rings can
behave quite differently from their bigger counterparts.


\subsubsection{Two-step or double switching}
The dominant switching process for magnetic rings is O-V-O switching
or O-T-O switching (see Fig. \ref{switching}(c)), since the
inevitable intrinsic asymmetry stimulates this process. A large
regime in the geometric phase diagram falls into this category. In
most cases the remanence state starting from saturation is an onion
state, and both transitions occur at negative field. When the
element is very thick, the remanence state can be a vortex state,
i.e. the first transition field is
positive.\cite{francePRL2001,triple} This behavior is found for
elements near the boundary between double and triple switching in
the switching phase diagrams. The mechanisms of the two steps are
discussed separately below.


For O-V or O-T, two possibilities exist, depending on whether it is
initiated by domain wall motion or nucleation.

(i) Nucleation and buckling: a strong buckling of the magnetization
happens first in one arm of the onion state where the magnetic
moments are antiparallel to the external field, followed by a
nucleation of a vortex passing through the
arm\cite{spinwave1,triple}. This process is not typical in two step
transitions. Rather, it happens in some shifted inner circle rings
or relatively wide rings. The transition field can be positive,
resulting in a vortex state at remanence. It is similar to the
triple switching-process, so it can also be regarded as a transient
process.

(ii) Nucleation free and domain wall motion: one wall is pinned more
than the other due to some sort of asymmetry, so that the other wall
moves towards it. When the two walls get close to each other, two
things could happen: (a) The annihilation process. This is possible
only if the ring is sufficiently wide (say larger than $10nm$); (b)
$360^o$ domain walls which are found to exist for narrow and small
rings recently\cite{RossPRB2003}. The states formed are twisted
states which have been discussed in the previous section. It is
interesting and enlightening to understand this process from the
topological point of view. As is shown in Sec.III, each of the
transverse domain walls consist of two half-vortices: one
$\omega=-1/2$ defect on the outer edge and one $\omega=+1/2$ defect
on the inner edge. The vortex state cannot form in arbitrarily thin
rings since the half vortices with the same winding number cannot
annihilate directly and they cannot move to bulk either. Only when
the vortex wall is also an energy minimum, one of the $\omega=+1/2$
edge defects can emit a $\omega=+1$ vortex into the bulk and
transform into a $\omega=-1/2$ defect. The emitted vortex will
travel to the other side and turn the $\omega=-1/2$ defect into a
$\omega=+1/2$ defect.\cite{top2005,saddle} Then the edge defects can
annihilate. Overall, the choice of the above processes depends
strongly on the ring width, but less so on $R_o$ and h (see Fig.
\ref{switchingp}). These two processes both happen for relatively
small widths.

The V-O transition is sometimes called a first magnetization curve.
In the half of the ring with magnetization antiparallel to magnetic
field $\vec{H}$, a vortex domain wall nucleates at the inner side of
the rim and passes through it perpendicular to $\vec{H}$. In the
meantime two transverse head-to-head domain-wall-like structures
appear next to the vortex and quickly propagate in opposite
directions, forming the final onion state. This process is mainly
shape dependent and has little to do with anisotropy, because it
depends on how easy it is to form a vortex wall on the edge, which
can be assisted by the edge imperfection (or roughness). As
expected, this process depends also on the local curvature of the
ring. Since the local curvature and edge roughness affect small
rings more, the distribution of the switching field will be larger
in these systems. However the magnitude of transition field $H_c$ is
insensitive to the radius. It increases with height
h\cite{KlauiJMMM2002} since thicker elements favor vortex
structures, and decreases with width as the vortex state is more
stable in thin rings. The nucleation requires a large twisting of
the spins, which is harder to achieve in thin rings. Simulations
tend to give higher $H_c$ in the absence of defects, but defects are
known to reduce $H_c$. This discrepancy is more severe for thin
rings.

The field distribution of O-V is typically larger than in the V-O
case and can mainly be attributed to the variation of intrinsic
defects among different rings. As temperature tends to wipe out the
effects due to defects, i.e. thermal fluctuation make it easier to
overcome the local energy barrier induced by defects, the
distribution becomes larger when the temperature goes down. This is
a little bit surprising at first glance. However, this conclusion
can be used to determine whether defects are responsible for the
field distribution. If the distribution depends strongly on
temperature, then defects are important. Generally speaking,
transitions involving nucleation processes are less prone to defects
and thermal fluctuation than processes involving domain wall or
vortex core motion.\cite{KlauiAPL2004} Since defects always exist,
simulations at 0K for perfect rings mimic more, in some sense, the
experimental situation at high temperature.


\subsubsection{Triple switching}

This process involve three intermediate steps: O-V-VC-O.
\cite{KlauiAPL2004,triple,triple2} It is similar to O-V-O except
that the vortex state does not deform into an onion state abruptly,
but nucleates a vortex core in one arm, and the core then moves
slowly to the outer rim. To some extent, it is similar to the
switching process of magnetic discs where the core in the vortex
state is shifted by a magnetic field and finally exits at the
boundary (see Fig. \ref{switching}(a)). To see this, Steiner
\textsl{et al} \cite{triple} have shown a sequence of hysteresis
curves with various inner radii $R_i$ from the disc limit ($R_i=0$).
The field distribution of the V-VC transition is affected little by
temperature, so it is affected little by defects. On the other hand
the VC-O is process affected strongly by temperature, so defects
play an important role here as cores can be pinned by defects. The
triple switching process only exists for thick and very wide rings.
Since these rings lack the advantages of rings mentioned before,
this switching process is less interesting and less studied.

Regarding the effects of anisotropy, it generally suppresses the
field distribution and increases the switching field. One curious
observation is that ring structures make the hard axis of the cubic
anisotropy into the global easy axis\cite{KlauiReview2003}.


\subsection{Circular field}

\begin{figure}[h]\begin{center}
\includegraphics[height=.3\textheight]{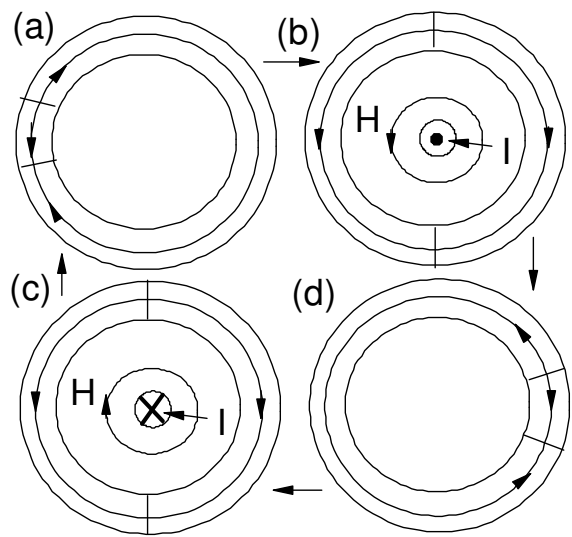}
\caption{\label{circ} Schematic diagram of twisted state reversal by
circular field. }
\end{center}\end{figure}
Circular fields are interesting because they can be naturally
generated by a perpendicular current through the disc center. This
design is especially suitable for rings, as there is a hole in the
center to deposit a nanodot to conduct current. As a result
current-induced switching can be achieved, which is essential for
data storage applications. At first, a circular field was proposed
to switch the vortex state into opposite
circulation\cite{cmuJAP2000}, but it requires a fairly large current
density and involves some fairly complicated transient states.
Recently, after the twisted states were discovered, a new scheme was
proposed.\cite{360domain,360domain2} As is shown in Fig. \ref{circ},
switching is adjusted between two twisted states with the $360^o$
domain wall located in the left and right rim (or any two positions
along a diameter). This switching requires low current and has a
very clean transition process which only involves a domain wall
movement. In addition, it is very fast.

\section{Applications}

Nanomagnets are featured in many applications. One of the most
important is in data storage devices. Three schemes have been
proposed\cite{cmuRev2007} for nanorings: (i) two onion states,
similar to the traditional oblong memory element, except that the
switching current is significantly less. However, the stray field is
an obstacle for high density storage; (ii) two vortex states with
different circulation, studied most intensively but requiring a
relatively high current density \cite{cmuJAP2000}; (iii) two twisted
states, most promising because of the low stray field and low
current. \cite{360domain}

For the first choice, Kl$\ddot{a}$ui \textsl{et al}
\cite{KlauiReview2003} have performed a dynamics switching study.
They used magnetic pulses to switch the magnetization. They found
that switching is only possible when $H*\Delta t\approx5*10^{-12}
T*s$ , keeping $H>20mT$ and $\Delta t>50ps$. The switching time is
T=0.4ns, so that the possible switching rate is about 1.25GHz
($=1/(2T)$).
%



Most attention has been focused on the second choice, for this one
utilizes all the benefits of nanorings. Manmade asymmetry offers a
way to control the vortex circulation. Several techniques are
available: shifted inner circle\cite{cmuPRL2006, controlPRL2008},
notches\cite{KlauiAPL2003}, and elongated shapes like
ellipses\cite{defect2,defect3}. Some
works\cite{cmuPRL2006,RossPRB2005} claim that even when there exists
asymmetry, the switching process is still a stochastic process due
to the thermal fluctuation. Some rings in an array may follow O-O
switching while other rings may follow O-V-O. Even for the same ring
the circulation may not be deterministic, so the strength of the
asymmetry needed is still an important open question.







Recently, increasing attention has been given to the third
possibility\cite{cmuRev2007,360domain,360domain2}, as it is
suggested that the twisted state can be switched very easily by a
small circular current and can achieve high speed. Experimental
evidence is needed to confirm the idea.


Another obstacle to application is that ring arrays always show a
wide distribution of switching fields. To make the application
realistic, the width of the distribution has to be reduced. The
distribution is attributed to both the interior and exterior
defects. Though the influence is qualitatively known, quantitative
analysis is still missing. Even though one would like to get rid of
defects as much as possible in most cases, they can also be used to
engineer the switching behavior. \cite{defect1,defect2}.



\section{Conclusions}
We have reviewed current results in the study of ferromagnetic
rings, with emphasis on switching processes. Ring structures feature
many advantages including being stable and insensitive to edge
imperfection and completely stray field free. Ring arrays give an
ultimate area storage density of about $0.25 Tbits/cm^2$. We covered
all possible states identified numerically and experimentally until
now, among which vortex, onion and twisted states are discussed in
detail, since they are suggested to be candidates for data storage.
There are three types of switching processes under uniform field,
depending on the geometric parameters of the ring. The double
switching, in particular, enjoys tremendous popularity. By
introducing engineered asymmetry, the circulation direction of the
vortex state can be easily controlled by a uniform field. The
mechanisms of these switching processes were explained carefully in
detail and topological argument were shown to be enlightening. On
the other hand, circular field switching is easy to generate, and is
suitable to switch the twisted state. Finally applications and
limitations were briefly discussed.\\



\textsc{\textbf{Acknowledgement:}} We would like to thank Noah
Jacobson and Yaqi Tao for useful discussions. Computing facilities
were generously provided by the University of Southern California
high-performance supercomputing center. We also acknowledge
financial support by the Department of Energy under grant
DE-FG02-05ER46240.\\

\bibliography{scale}

\end{document}